\documentclass[aps,prd,onecolumn,groupedaddress,nofootinbib,preprintnumbers,superscriptaddress]{revtex4-2}  
\usepackage{epsfig}
\usepackage[top=75pt,bottom=75pt,left=50pt,right=50pt]{geometry}
\usepackage{enumitem}
\usepackage{caption}
\usepackage{amsmath}
\usepackage{amssymb}
\usepackage{bbm} 
\usepackage{upgreek} 
\usepackage{graphicx}
\usepackage{booktabs}
\usepackage{epstopdf}
\usepackage{gensymb}
\usepackage[colorlinks=true,citecolor=blue,linkcolor=blue]{hyperref}
\usepackage[capitalise,noabbrev]{cleveref}
\usepackage{bm} 

\usepackage{tikz,xcolor}
\hypersetup{
  unicode    = false,
  pdfcreator = {RevTeX},
  colorlinks = true,
  linkcolor  = red,
  citecolor  = blue,
  filecolor  = black,
  urlcolor   = blue
}
\usepackage[utf8]{inputenc}

\definecolor{lime}{HTML}{A6CE39}
\DeclareRobustCommand{\orcidicon}{\hspace{-4pt}
	\begin{tikzpicture}
		\draw[lime, fill=lime] (0,0) 
		circle [radius=0.16] 
		node[white] {\hspace{0.1mm}{\fontfamily{qag}\selectfont \tiny ID}};
		\draw[white, fill=white] (-0.07,0.1) 
		circle [radius=0.01];
	\end{tikzpicture}
	\hspace{-3.2mm}
}

\foreach \x in {A, ..., Z}{\expandafter\xdef\csname 
	orcid\x\endcsname{\noexpand\href{https://orcid.org/\csname orcidauthor\x\endcsname}
		{\noexpand\orcidicon}}
}

\begin{document}

\title{A closer look at the LZ 248 keV event through the lens of\\ cosmic-ray boosted dark matter}
 
\author{Bhavesh Chauhan \orcidA}
\affiliation{Department of Physics, Birla Institute of Technology and Science, Pilani 333031, India}

\author{Soham Sahasrabuddhe \orcidC}
\author{Manibrata Sen \orcidB}

\affiliation{Department of Physics, Indian Institute of Technology Bombay, Powai, Mumbai 400076, India}

\date{\today}

\begin{abstract}
We investigate whether cosmic ray boosted dark matter (CRDM) can explain the 248 keV nuclear recoil event observed by the LUX-ZEPLIN (LZ) collaboration. We consider spin-dependent interactions, mediated by a pseudoscalar, and incorporate relativistic kinematics in CRDM production, propagation, and detection. We find that spin-dependent interactions can produce sufficient number of nuclear recoils in the observed region of interest, without populating events at lower energies, which have not been observed. However, for a benchmark pseudoscalar interaction corresponding to the non-relativistic effective operator $\mathcal{O}_6$, the coupling required to produce the observed event corresponds to an energy scale comparable to the momentum transfer scales, thus making contact interaction approach challenging. Instead, we find that a pseudoscalar mediator with mass $\sim$300\,MeV and $\mathcal{O}(1)$ couplings can provide a promising solution to this unexplained event.  
\end{abstract}

\maketitle

\section{Introduction}
\label{sec:intro}

The LUX-ZEPLIN (LZ) experiment has recently extended its nuclear-recoil search to energies of approximately $270~\mathrm{keV}$, reporting a single event with reconstructed energy, $E_{\rm NR} = 248 \pm 23~(\mathrm{stat}) \pm 23~(\mathrm{sys})~\mathrm{keV}$, in a $2.84~\mathrm{t\,yr}$ exposure, corresponding to $4.71~\mathrm{t}$ of fiducial mass and $220$ live days~\cite{LZ2026}. The event occurs in a region of low expected background and corresponds to a global $2.6\sigma$ tension with the background-only hypothesis after accounting for the look-elsewhere effect, with a maximum local significance of $3.4\sigma$. While the observation does not constitute evidence for dark matter (DM), its unusually high recoil energy raises a simple phenomenological question: what classes of dark matter interactions can produce a recoil spectrum concentrated near $250~\mathrm{keV}$?

The LZ collaboration's own interpretation uses non-relativistic effective field theory (NREFT) operators and inelastic DM, and favours weakly interacting massive particle (WIMP) masses above $200~\mathrm{GeV}$, where the recoil spectrum extends far enough in energy for the event bin to carry an appreciable fraction of the rate. Following the observation of the event, several interpretations have been proposed. These include inelastic scenarios~\cite{Nagata:2026pbj, An:2026pkc, yin2026, bandyopadhyay2026, fan2026, jeesun2026, rodd2026, Ge:2026xax, Xing:2026civ, Borah:2026ris, Ghosh:2026txe, Das:2026buc, Lian:2026hpm, Ahmed:2026kan, He:2026idw, Nguyen:2026lui, Chatterjee:2026scv, Fan:2026hzw, Qi:2026vyp, Kumar:2026lgi, Baer:2026fpy, Yuan:2026djt, Cheung:2026byg, Lee:2026jxl, Asadi:2026iot, Langhoff:2026ujr, Okada:2026eol, Ahmed:2026qjg, Du:2026lpa, Borah:2026zwf, Bisal:2026khf, Bose:2026ndd, Wang:2026ytg,DiMauro:2026dqp, Das:2026uyy, Lee:2026wof, Dent:2026bji, Gu:2026vto,deLima:2026shq, Smirnov:2026aqk, Du:2026guj, Pospelov:2026ewn,Visinelli:2026kgt, Nomura:2026qyq, Yamashita:2026ump, Su:2026rwz, Freese:2026sga, Wu:2026nhi}, boosted dark matter~\cite{Bell_2024, Mahapatra:2026glu, Heikinheimo:2026kwp, Alhazmi:2026efz, Kannike:2026qyl, Liang:2026coz}, and other interesting phenemenological studies~\cite{dimauro2026,mccabe2026, Lueiza-Colipi:2026gij, Arcadi:2026kev, Barman:2026omh, Palmisano:2026kuj, Uttayarat:2026isp, DiMauro:2026ymt, Okada:2026upm, Lee:2026zbr, Chattaraj:2026fxn, He:2026hqz, Egorov:2026dpr, Bamwidhi:2026vdu, Elahi:2026vlm,Zhu:2026dag,Aghaie:2026vsu, Lee:2026xxh, Khan:2026nwp, Yang:2026wpb, Kotlarski:2026pep,Unwin:2026rdp}. 

A high energy nuclear recoil, however, does not necessarily require heavy WIMP DM. A light DM particle can acquire relativistic kinetic energy through scattering with galactic cosmic rays, producing a population of cosmic-ray boosted dark matter (CRDM) particles~\cite{Cappiello:2018hsu, Bringmann_2019}. 
Such particles can transfer substantially more energy to a nucleus than is possible for nonrelativistic halo DM of the same mass, opening a qualitatively different route to the LZ event. CRDM is therefore particularly interesting for sub-GeV DM, for which conventional nuclear-recoil searches are normally limited by kinematics \cite{Bardhan_2023, Cappiello:2024acu, Bell_2024, Herbermann_2024, Jeesun:2026lro}.

In this work, we investigate whether CRDM can account for the spectral properties of the LZ event. We consider only spin-dependent (SD) interactions and consistently include relativistic kinematics in CRDM production, propagation, and detection. A key point is that the CRDM is relativistic, so the usual NREFT methods cannot be applied directly to these scattering processes. Instead, for each of the SD nuclear operators considered, we employ the corresponding Lorentz-invariant interaction, following~\cite{Del_Nobile_2018} and retain the exact relativistic DM current. We employ a semi-relativistic (SR) approach, where the DM current is relativistic, whereas the nucleonic current is treated in the NREFT formalism. Such methods have been used in the context of lepton-nucleus scatterings, but have not been used for cosmic-ray boosted DM. This is what distinguishes our work from previous studies on CRDM where nuclear responses have been modeled via a form-factor. The resultant CRDM is propagated through the Earth using the continuous slowing down approximation, and the recoil spectrum in LZ is calculated using the corresponding nuclear response functions.

We find a qualitative difference between SI and SD interactions. For coherent SI scattering, following the analysis of Bringmann and Pospelov \cite{Bringmann_2019}, we find that fitting the LZ event leads to substantial overproduction of lower-energy recoils, and hence these are inefficient in explaining the results. In contrast, SD interactions can produce spectra concentrated near the observed energy, because of the combination of momentum dependence and the finite-momentum spin responses of the xenon nuclei. As a benchmark case, we demonstrate our results by considering the pseudoscalar interaction  mediated by the effective operator of the form $ \left( \overline{\chi} i\gamma_5 \chi \right) \left( \overline{N} i\gamma_5 N \right)$. 

We also compare results obtained under the  the non-relativistic (NR), form-factor (FF), and semi-relativistic (SR) prescriptions for the DM-nuclear scattering. Although the NR and SR approaches agree, the FF prescription differs significantly at the momentum transfer relevant to the LZ region of interest. In particular, the FF has a zero near 250\,{\rm keV} for xenon, leading to a significant under-estimation of the predicted rate in the event window. This has to be considered carefully when interpreting high energy recoil events in these direct detection experiments.

Despite the favourable spectral shape of the SD recoil spectrum, we find that the pseudoscalar interactions cannot account for the LZ event within the assumptions of the analysis. Requiring that the coupling produces the observed event in the LZ window of interest leads to the energy scale of the effective operator to lie in the 40-90 MeV window. This is far below the momentum transfers involved both in the scattering and detection process, thereby invalidating the effective field theory treatment.We therefore conclude that CRDM with the contact pseudoscalar interaction cannot successfully explain the LZ event. An obvious way out involving boosting DM by scattering through sub-GeV mediators can help alleviate this, as we will discuss below.

Our paper is organised as follows. Sec.~\ref{sec:methods} describes the interaction models and operators considered, followed by the CRDM production mechanism, attenuation in the Earth, and the event-rate calculation at LZ. In Sec.~\ref{sec:results}, we present the main results of our analysis and assess whether CRDM can account for the observed LZ event. Finally, Sec.~\ref{sec:summary} summarises our findings and conclusions.

\section{Model description and methods}\label{sec:methods}

In this section, we revisit the production of CRDM following the standard formalism \cite{Cappiello:2018hsu, Bringmann_2019}, and discuss the inputs used in our analysis. We estimate the flux of CRDM at the LZ detector depth that survives passing through the Earth, and use this attenuated flux to calculate the event rate of observable nuclear recoils. All three pieces (production, attenuation, detection) require the differential cross-section for DM scattering off nuclei, which we obtain from a phenomenological model. At the detector, the CRDM is relativistic, and consequently, the usual NREFT framework cannot be applied. We discuss two approaches that were adopted in this work to estimate the differential cross-sections. 

We take the local DM density to be $\rho_\chi = 0.3~\mathrm{GeV\,cm^{-3}}$ \cite{Catena:2011kv} and neglect the halo DM velocity, $v_{\rm halo}\sim10^{-3}$, compared with the relativistic cosmic-ray velocities. The DM is assumed to be a spin-$1/2$ Dirac fermion. Throughout, $T_\chi$ denotes the DM kinetic energy, $T_i = E_i - m_i$ denotes the kinetic energy of the cosmic-ray particle, and $E_R$ denotes the kinetic energy of the recoiling target nuclei of mass $m_T$ and spin $j_T$. As usual, $Q^2 = -q^2>0$ denotes the invariant momentum transfer. 

\subsection{Interaction Model}

We adopt a phenomenological model of DM-nucleon interactions, in which we consider a Lorentz-covariant interaction that generates a SD interaction in the non-relativistic limit. From the several choices listed in \cite{Anand_2014}, we consider a pseudo-scalar interaction as an example. A detailed study of all interactions is left as future work. In this work, the DM-nucleon interaction Lagrangian is assumed to be, 
\begin{equation}\label{eq:int_lag}
    - \mathcal{L}_{\rm int} = \sum_{N=p,n} G^{N}\, \left( \overline{\chi} i\gamma_5 \chi \right) \left( \overline{N} i\gamma_5 N \right), 
\end{equation}
where $N = (p,n)$ denotes the nucleon. We consider two benchmark scenarios: (a) isoscalar couplings (i.e., $G^p = G^n$), and (b) proton-only coupling (i.e., $G^p \neq 0,  =G^n = 0$). In the NREFT framework, the couplings are written in the nuclear isospin basis, i.e., $G^0 = (G^p + G^n)/2$ and $G^1 = (G^p - G^n)/2$. 

\subsection{Production of cosmic-ray boosted dark matter}

A halo DM particle can undergo an elastic scattering with a cosmic-ray particle ($\text{CR}_i + \chi \rightarrow \text{CR}_i + \chi$) and recoil with a relativistic boost. The flux of these boosted DM particles at the top of Earth's atmosphere is given as \cite{Bringmann_2019}, 
\begin{equation}
    \frac{d\Phi_\chi}{d T_\chi}
    =D_{\rm eff}\frac{\rho_\chi}{m_\chi}
    \sum_i\int_{T_i^{\min}(T_\chi)}^{\infty}\!\!d T_i\;
    \frac{d\Phi_i^{\rm LIS}}{d T_i}\,
    \frac{d\sigma_{\chi i}}{d T_\chi},
    \label{eq:flux}
\end{equation}
where $i$ indexes the dominant species of cosmic rays (i $\in \{p, \mathrm{He}, \mathrm{C}, \mathrm{O}, \mathrm{Fe}\}$). In \eqref{eq:flux}, $D_{\rm eff}$ is the line-of-sight integral over the halo density which multiplies the local production rate per unit volume. For a Navarro-Frenk-White profile~\cite{Navarro_1997} and a homogeneous CR density, integrating out to $1~\mathrm{kpc}$ ($10~\mathrm{kpc}$) gives $D_{\rm eff} = 0.997~\mathrm{kpc}$ ($8.02~\mathrm{kpc}$)~\cite{Bringmann_2019}, and a calculation with spatially dependent CR densities finds $D_{\rm eff} \simeq 9~\mathrm{kpc}$ for CRDM kinetic energies above $\sim 1~\mathrm{GeV}$~\cite{Xia_2022}. Following Ref.~\cite{maity2024}, we use $D_{\rm eff} = 10~\mathrm{kpc}$ throughout. For the local interstellar spectrum (LIS) of cosmic-rays, we used the Gaisser-Stanev-Tilav global fit~\cite{gaisser2013}.

In Ref. \cite{Bringmann_2019}, the differential cross-section for the scattering was assumed to be isotropic and written as, 
\begin{equation}
    \frac{d\sigma_{\chi i}}{d T_\chi}\Bigg|_{\rm BP} = \frac{\sigma_{\chi i}}{T_{\chi}^{\rm max}(T_i)} \times F_i^2 \left( Q^2 = 2 m_\chi T_\chi\right),
\end{equation}
where $F_i(Q^2)$ is the form-factor and $T_\chi^{\rm max}(T_i)$ is the maximum kinetic energy that leads to a recoil with kinetic energy $T_i$. The expression for $T^{\rm max}$ and $T^{\rm min}$ are adapted from Ref.\,\cite{Bringmann_2019}. 

For the phenomenological Lagrangian considered in this work, we evaluate the differential cross-section, which brings in momentum dependence. The standard NREFT, which provides the nuclear response functions but assumes non-relativistic DM, cannot be applied directly to CRDM. To address this, we consider two approaches: (a) a Lorentz-covariant formalism with a form-factor (FF) for the nucleus, and (b) a semi-relativistic (SR) approach where the DM current is treated in a Lorentz-covariant formalism, and the nuclear current is treated in the NREFT formalism. The first approach is simple to execute but depends on modelling the form factor. The second approach is advantageous because it utilises the nuclear response functions. 

Upon taking the non-relativistic limit, the interaction Hamiltonian for \eqref{eq:int_lag} at leading order in $|\vec{q}|/m_N$ is given by 
\begin{equation}\label{eq:int_ham}
    \mathcal{H}_{\rm int} =  \sum_{\tau=0,1} G_{}^\tau \,\left( \frac{m_N}{m_\chi} \right) \left( \vec{S}_\chi \cdot \frac{\vec{q}}{m_N} \right)\left( \vec{S}_N \cdot \frac{\vec{q}}{m_N} \right),
\end{equation}
where $m_N$ is the nucleon mass, $m_\chi$ is the mass of DM, $S_{\chi}$ is the spin of the DM, and $S_N$ is the spin of the nucleon. On comparing with the basis from Ref.\,\cite{Anand_2014}, we see that $c_6^\tau = G_{}^\tau\,\left(m_N/m_\chi \right)$, and $c_i=0$ for all other terms. The differential cross-section for a scattering between a \emph{non-relativistic} DM with velocity $v$ and a target nucleus of mass $m_T$ at rest is given by,
\begin{equation}\label{eq:diffxs_NR}
    \dfrac{d\sigma_{\chi T}}{d E_R}{\Bigg |}_{\rm NR} =  \frac{m_T |\vec q|^4 }{32 \pi m_N^2 m_\chi^2 v^2} \,\left( \frac{4\pi}{2 j_T + 1} \right) \, \left[\sum_{\tau, \tau^\prime} G_{}^{\tau} G_{}^{\tau^\prime} W_{ T}^{\tau \tau^\prime} \right],
\end{equation}
where $W_{T}^{\tau\tau'}$ is the nuclear response that depends on $|q|$. For the case of isoscalar interactions, the term inside the square brackets of \eqref{eq:diffxs_NR} simplifies to  
$ (G_{}^{0})^2 W_{T}^{00}$, and for the case of proton-only it simplifies to $(G_{}^{0})^2 ( W_{T}^{00} + W_{T}^{01} + W_{T}^{10} + W_{T}^{11})$. We have used \texttt{WimPyDD}\,\cite{Jeong_2022} for evaluating the nuclear response functions.

In the FF approach, the DM-nucleus cross-section is a product of the DM-nucleon cross-section of \eqref{eq:int_lag} and a form factor $F(q)$, which models the distribution of nucleons inside the nucleus~\cite{LS}. For the pseudo-scalar interaction, the nucleus couples through its total proton and neutron spins, and the differential cross-section for a DM particle of momentum $p_\chi$ scattering off a nucleus at rest is~\cite{LS}
\begin{equation}\label{eq:diffxs_FF}
    \dfrac{d\sigma_{\chi T}}{d E_R}{\Bigg |}_{\rm FF} = \frac{m_T\, Q^2\, |\vec q|^2}{32 \pi\, m_N^2\, p_\chi^2}\; \frac{4}{3}\,\frac{j_T+1}{j_T}\, \Big( G_{}^p \langle S_p\rangle + G_{}^n \langle S_n\rangle \Big)^2 \, F^2(q),
\end{equation}
where $p_\chi^2 = T_\chi(T_\chi + 2 m_\chi)$, $Q^2 = 2 m_T E_R$, $|\vec q|^2 = E_R (E_R + 2 m_T)$, and $\langle S_{p,n}\rangle$ are the expectation values of the total proton and neutron spin in the nuclear ground state. Following Fitzpatrick \emph{et al.} \cite{Fitzpatrick_2013}, we use $\langle S_{p}\rangle = 0.007 $ and $\langle S_{n}\rangle= 0.248$ for $^{\rm 129}$Xe, and $\langle S_{p}\rangle = -0.005 $ and $\langle S_{n}\rangle= -0.199$ for $^{\rm 131}$Xe\footnote{Alternatively, $\langle S_{p} \rangle = 0.010 $ and $\langle S_{n}\rangle= 0.329$ for $^{\rm 129}$Xe, and $\langle S_{p}\rangle = -0.009 $ and $\langle S_{n}\rangle= -0.272$ for $^{\rm 131}$Xe}. For $F(q)$, we instead use the thin-shell form-factor of Lewin and Smith~\cite{LS},
\begin{equation}\label{eq:LS_ff}
    F^2(q) = \begin{cases}
        j_0^2(q r_n) & q r_n < 2.55 \;\; \text{or} \;\; q r_n > 4.5, \\[2pt]
        0.047 & 2.55 \le q r_n \le 4.5,
    \end{cases}
\end{equation}
where $r_n = 1.0\, A^{1/3}~\mathrm{fm}$ and $j_0$ is the spherical Bessel function.

In the SR approach, we retain the exact relativistic DM current and treat only the nucleus non-relativistically, so that the shell-model nuclear response functions of Ref.~\cite{Anand_2014} can be used. This method has previously been applied in the context of neutrino nucleus interactions \cite{Altmannshofer:2018xyo}. For the contact interaction of \eqref{eq:int_lag}, the amplitude for $\chi(p) + T(k) \to \chi(p^\prime) + T(k^\prime)$ factorises into a DM matrix element and a nuclear matrix element~\cite{Anand_2014}. The differential cross-section for a relativistic DM particle of momentum $p_\chi$ scattering off a nucleus at rest is
\begin{equation}\label{eq:diffxs_SR}
    \dfrac{d\sigma_{\chi T}}{d E_R}{\Bigg |}_{\rm SR} = \frac{m_T\, Q^2\, |\vec q|^2}{32 \pi\, m_N^2\, p_\chi^2}\, \left( \frac{4\pi}{2 j_T + 1} \right)\, \left[ \sum_{\tau, \tau^\prime} G_{}^{\tau} G_{}^{\tau^\prime} W_{T}^{\tau \tau^\prime}(y) \right]F^2(Q^2),
\end{equation}
for $F(Q^2)$ is form-factor for a nucleon which we take to be of the dipole form with $\Lambda=770$\,MeV. This form-factor accounts for the fact the nuclear-matrix elements in Ref. \cite{Anand_2014} are evaluated assuming nucleons to be point particles. However, at the momentum transfer relevant for the LZ event, this approximation starts to break down. In the non-relativistic limit, \eqref{eq:diffxs_SR} reduces to the NREFT result of \eqref{eq:diffxs_NR}.

\subsection{Attenuation of boosted dark matter due to Earth}

The LZ detector is located at a depth $d = 1.478\,\mathrm{km}$ at the Sanford Underground Research Facility \cite{Akerib_2020}. Before reaching the detector, CRDM particles traverse a distance between $d$ and $2R_E-d$ through the Earth, depending on their direction of incidence. During this propagation, they lose energy through scattering with nuclei in the Earth, modifying the incident CRDM spectrum. We describe this attenuation within the continuous slowing down approximation (CSDA) ~\cite{Bringmann_2019,maity2024,Herbermann_2024,Das:2024ghw}, in which the DM kinetic energy evolves deterministically according to
\begin{equation}\label{eq:dTdx}
    \frac{dT_\chi}{dx}(x)
    = -\sum_T n_T \int_0^{E_R^{\max}(T_\chi)} dE_R\, E_R\, \frac{d\sigma_{\chi T}}{dE_R}
    \equiv -S(T_\chi).
\end{equation}
Here, the sum runs over the relevant target nuclides, which differ between spin-independent and spin-dependent interactions, $n_T$ is the number density of the target nuclide $T$, and $d\sigma_{\chi T}/dE_R$ is the corresponding differential scattering cross-section. Equation~\eqref{eq:dTdx} can be solved numerically to obtain the DM kinetic energy $T_\chi(x)$ for an incident kinetic energy $T_\chi^0$. We instead invert this relation to determine the required initial energy $T_\chi^0(T_\chi,x)$ for a particle to reach the detector with kinetic energy $T_\chi$ after traversing a distance $x$.

Following Refs.~\cite{Bringmann_2019, Bell_2024}, we consider only elastic scattering in the attenuation calculation. Inelastic and quasi-elastic channels can become relevant for sufficiently large interaction strengths~\cite{Gu:2026gec, Bell_2024}, but are not included in the present analysis. In contrast to the isotropic assumption of Ref.~\cite{Bringmann_2019}, we retain the full dependence of $d\sigma_{\chi T}/dE_R$ on both $E_R$ and $T_\chi$ in our analysis.
We model the Earth as continental crust with density $\rho_c = 2.7~\mathrm{g\,cm^{-3}}$~\cite{maity2024}, composed of its eleven most abundant elements. For spin-independent interactions, we include coherent elastic scattering from all of these elements. For spin-dependent interactions, only nuclides with non-zero spin contribute. In this work, we include $^1$H, $^{23}$Na and $^{27}$Al, for which shell-model spin responses have been computed using \texttt{WimPyDD}\,\cite{Jeong_2022}.

Along each path, we assume that the stopping medium has a uniform composition, so $S$ depends only on position, through the density. The energy-loss equation can therefore be integrated once for the assumed crust composition. A particle that reaches the detector with kinetic energy $T_\chi^z$ after traversing an equivalent path length $\ell$ at crust density has an initial kinetic energy $T_\chi^0$ determined by
\begin{equation}\label{eq:range}
    \int_{T_\chi^z}^{T_\chi^0} \frac{dT}{S(T)} = \ell ,
\end{equation}
where $S$ is evaluated at the crust density. Differentiating equation~\eqref{eq:range} at fixed $\ell$ gives the Jacobian
\[
\frac{dT_\chi^0}{dT_\chi^z} = \frac{S(T_\chi^0)}{S(T_\chi^z)}.
\]
The attenuated flux at the detector is then~\cite{Das:2024ghw}
\begin{equation}\label{eq:fluxz}
    \frac{d\Phi_\chi}{dT_\chi}\Bigg|_{\rm LZ} = \int \frac{d\Omega}{4\pi}\, \frac{d\Phi_\chi}{dT_\chi}\bigg|_{T_\chi^0}\frac{dT_\chi^0}{dT_\chi^z}
    = \frac{1}{2}\int_{-1}^{1} d\cos\theta\; \frac{d\Phi_\chi}{dT_\chi}\bigg|_{T_\chi^0(T_\chi^z,\,\ell(\theta))}\, \frac{S\big(T_\chi^0\big)}{S\big(T_\chi^z\big)},
\end{equation}
where $d\Phi_\chi/dT_\chi$ is the isotropic flux of equation~\eqref{eq:flux}, $\theta$ is the zenith angle of arrival and $\ell(\theta)$ is the corresponding path length through the Earth. Directions for which equation~\eqref{eq:range} has no solution do not contribute.

\subsection{Event Rate at LZ}

The differential spectrum of nuclear recoils is given by
\begin{equation}\label{eq:rate}
    \frac{d \mathcal{N}}{dE_R} = \sum_T N_T \, t_{\rm live} \int_{T_\chi^{\min}(E_R)}^{\infty} dT_\chi\, \epsilon(E_R) \frac{d\Phi_\chi}{dT_\chi}\Bigg{|}_{\rm LZ}\, \frac{d\sigma_{\chi T}}{dE_R}(T_\chi, E_R),
\end{equation}
where the sum runs over xenon isotopes $T$, $N_T$ is the number of target isotopes, the nuclear-recoil detection efficiency is $\epsilon$ (Fig.~S2 of Ref.~\cite{LZ2026}), and we consider the exposure of $2.84\,\mathrm{t\,yr}$ which is the exposure for the analysis that reported the 248 keV event\,\cite{LZ2026}. We do not apply the detector energy resolution, so $E_R$ in equation\,\eqref{eq:rate} is the true recoil energy. For SD interactions, only $^{129}$Xe ($j_T = 1/2$) and $^{131}$Xe ($j_T = 3/2$) contribute as the other isotopes have $j_T = 0$.

\section{Results and Discussion}
\label{sec:results}

\begin{figure}[t]
\centering
\includegraphics[width=0.45\linewidth]{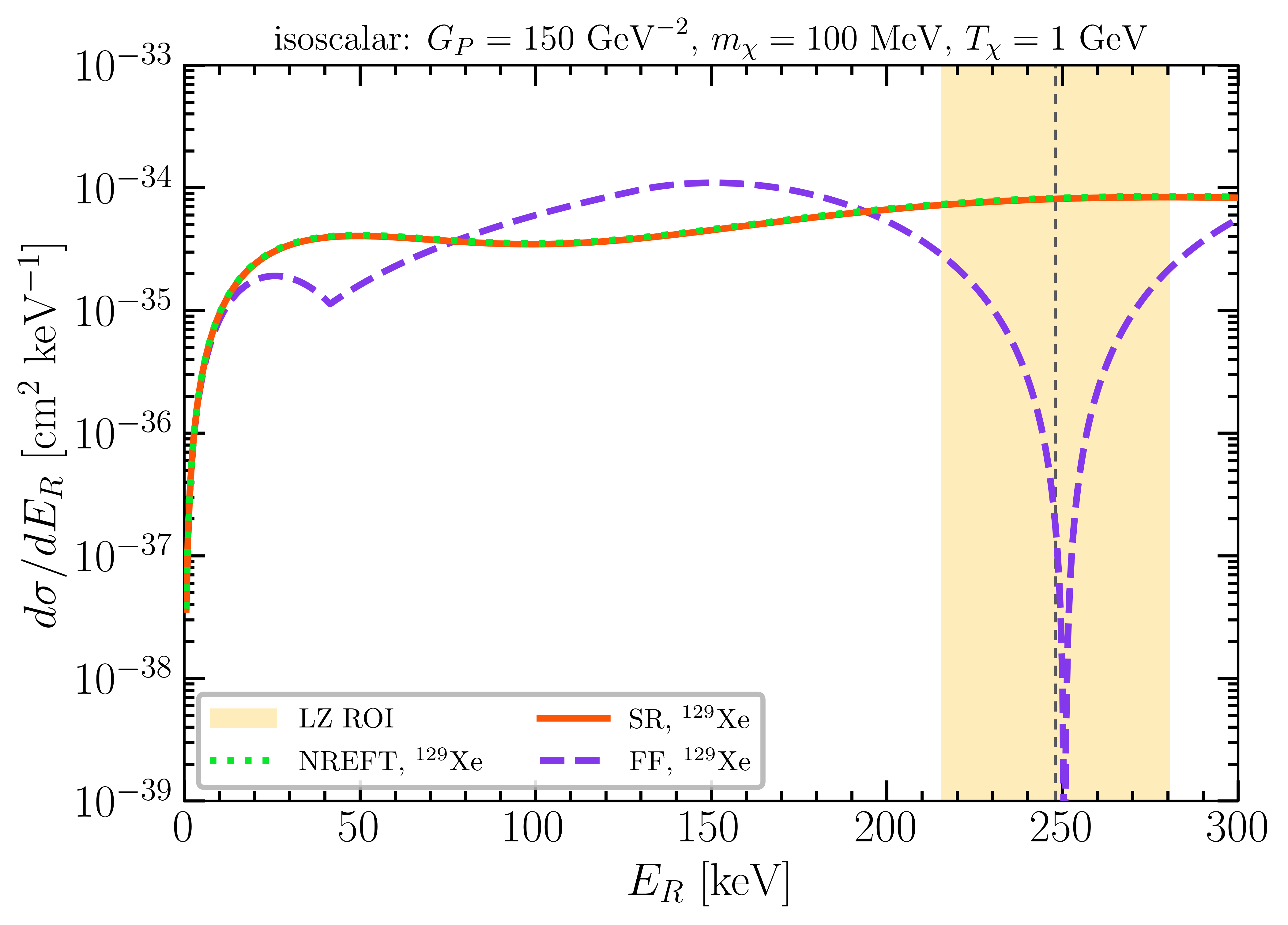}
\includegraphics[width=0.45\linewidth]{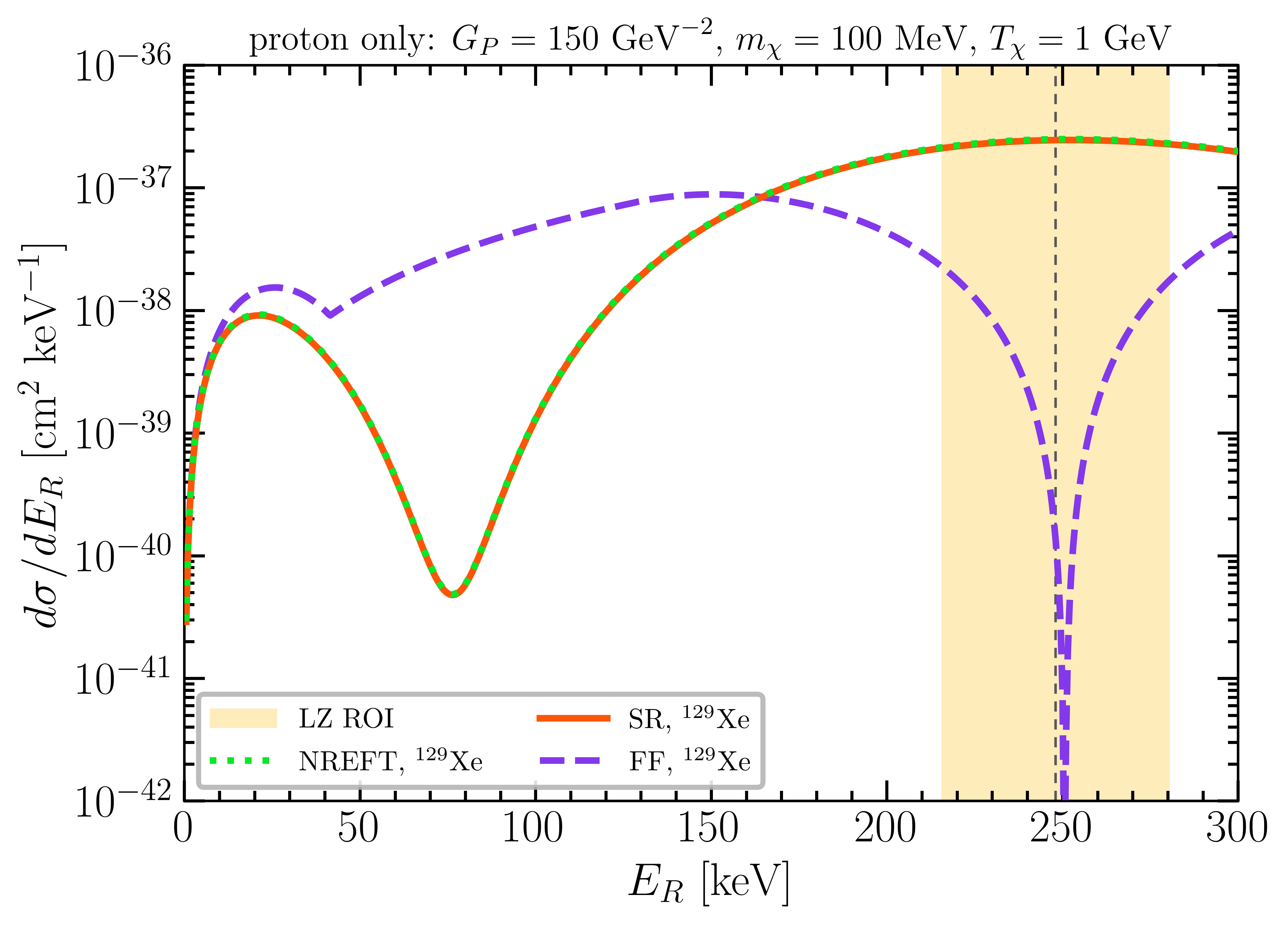}
\caption{Differential cross-section on $^{129}$Xe at $T_\chi = 1~\mathrm{GeV}$ for isoscalar coupling (\emph{left}) and proton-only coupling (\emph{right}).}
\label{fig:xsec}
\end{figure}

We fix $m_\chi = 100~\mathrm{MeV}$ throughout as a benchmark DM mass. Since only cosmic-ray protons produce the flux, the two cases (i.e., isoscalar and proton-only coupling) have the same flux before attenuation. The difference primarily arises from the interactions inside the Earth and in the detector. In figure~\ref{fig:xsec} we show $d\sigma/dE_R$ on $^{129}$Xe for the three prescriptions at $T_\chi = 1~\mathrm{GeV}$. The FF and SR results coincide as $q\to0$, and they start to differ as soon as $q r_n \sim 1$. The thin-shell form factor has a plateau for $2.55 < qr_n < 4.5$ and a zero at $qr_n = 2\pi$, which for xenon sits at $E_R \simeq 250~\mathrm{keV}$, right at the LZ event. The shell-model response has no zero in this range. A FF prescription would therefore underestimate the event rate exactly where LZ has observed the event. The differential cross-sections in the NREFT and SR approach are equal up to $(1 + Q^2/4m_T^2)$, which is $(1 + 10^{-6})$ for xenon and hence the curves lie on top of each other.

We keep only the flux arriving from above, where the path in rock is $1.5$ to $137~\mathrm{km}$, and treat the energy loss in using CSDA. The energy loss is almost entirely on hydrogen, and there the NREFT and SR cross-sections do not overlap even though they coincide for xenon. For hydrogen, the NREFT stopping power is larger, and the predicted rates differ by a factor of $\mathcal{O}(1)$ for this reason. The bigger issue is that the CSDA is a rough prescription for attenuation, and the correct approach requires detailed Monte Carlo simulations \cite{Xia_2022}. The impact of detailed simulations and analytical estimates can be seen clearly in Ref.\,\cite{LZ:2025iaw}. Moreover, we have not included the inelastic scattering and quasi-elastic-like scatterings that will have a significant contribution to attenuation at these energy scales. Hence, for simplicity, we only consider the CRDM from above and ignore the flux from below, which also depends on the poorly known hydrogen content of the core. Our rates and one-event couplings should therefore be read as estimates, and a detailed collision-by-collision transport calculation is left as a future exercise.

\begin{figure}[t]
\centering
\includegraphics[width=0.8\linewidth]{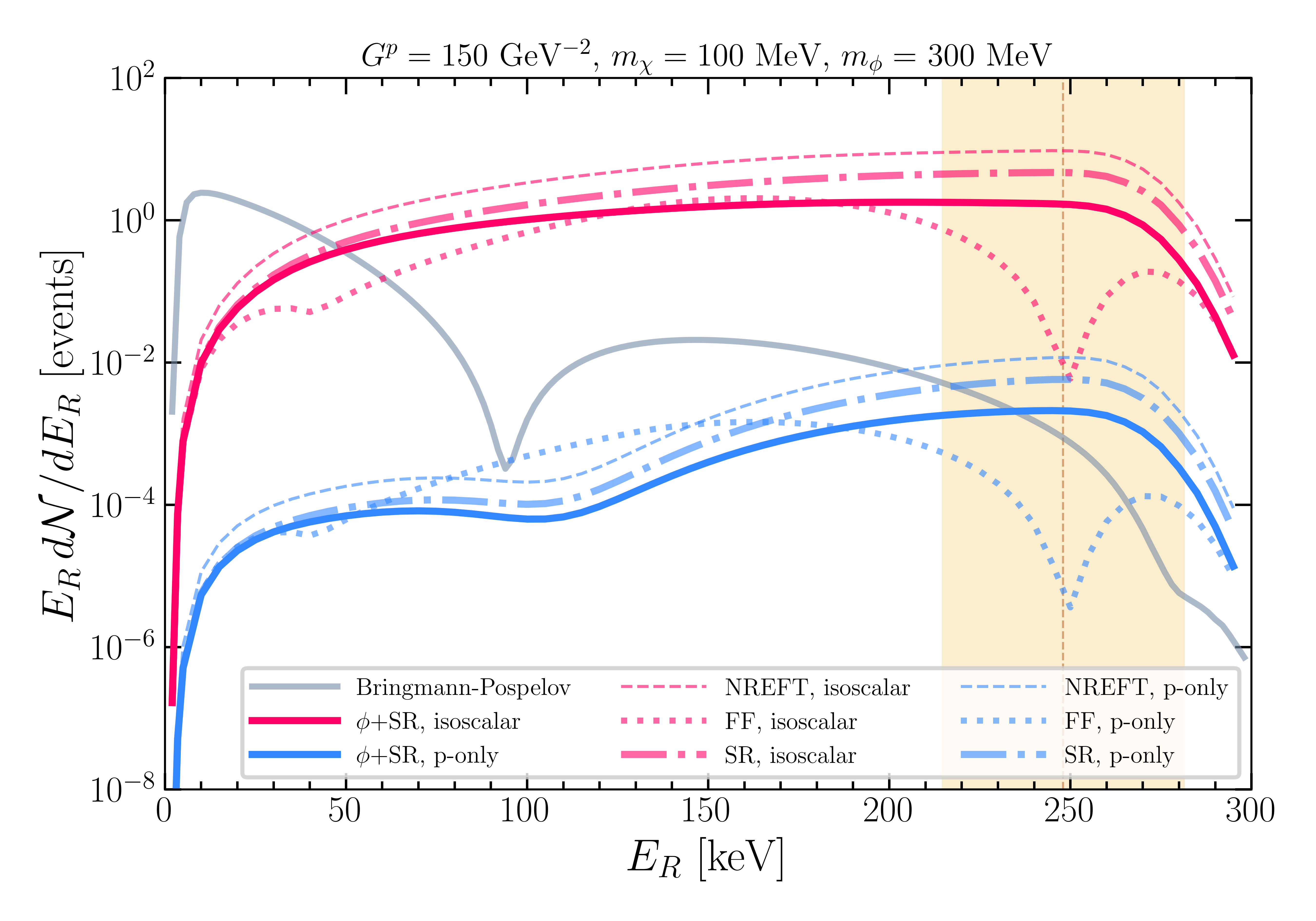}
\caption{Nuclear-recoil spectra in LZ for $2.84~\mathrm{t\,yr}$, efficiency applied, flux from above. The band is the LZ event, $248 \pm 33~\mathrm{keV}$.}
\label{fig:rate}
\end{figure}

In Figure~\ref{fig:rate} we show the nuclear recoil spectra for the LZ detector originating from CRDM with interactions as modeled in this work. We find that the SI spectrum falls steeply while the SD spectrum, for both SR and NREFT prescription, is nearly flat up to the event bin. As a result, the CRDM  cross-section that yields one event in the observed energy range would predict a large number of events at lower recoils, which should have been observed by LZ or XENON. Consequently, the SD interactions, or more generally, momentum-dependent interactions, allow a more plausible interpretation of the LZ event. This conclusion was also reached in Ref.\,\cite{Heikinheimo:2026kwp}. Moreover, within the model we find that the proton-only interaction model lead to lesser events at lower energies when compared with isoscalar interaction model. This feature arises primarily due to energy-dependent cancellations in the nuclear response functions. The proton-only case is additionally suppressed by $\langle S_p\rangle^2/\langle S_n\rangle^2 \sim 10^{-3}$ on the odd-neutron xenon isotopes. The NREFT and SR spectra agree upto few$\,\%$, and the FF spectrum has additional suppression due to the minima being in the event region of interest. This was also pointed out in Ref.\,\cite{Kannike:2026qyl}. 

We find that, in order to explain the observed event, one requires a contact interaction strength of $\mathcal{O}(100)\,\text{GeV}^{-2}$. Assuming $\mathcal{O}(1)$ coupling strength, this requires a mediator of mass $\mathcal{O}(100)\,\text{MeV}$. Since $q\sim250\,$MeV, the contact operator approximation is not the best description and one needs the complete propagator treatment to obtain the cross-section. Assuming that the contact operator in \eqref{eq:int_lag} is generated by integrating out a pseudoscalar $\phi$ of mass $m_\phi$ with $g_\chi$($g_N$) as a coupling to DM (nucleon), the effective interaction strength $G^{N}$ can be replaced by $g_\chi g_N/(Q^2 + m_\phi^2)$
and thus the cross-section is given as,
\begin{equation}\label{eq:diffxs_mphi}
    \dfrac{d\sigma_{\chi T}}{d E_R}{\Bigg |}_{\rm \phi+SR} = \frac{m_T\, Q^2\, |\vec q|^2}{32 \pi\, m_N^2\, p_\chi^2}\, \left( \frac{m_\phi^2}{Q^2 + m_\phi^2}\right)^2\,\left( \frac{4\pi}{2 j_T + 1} \right)\, \left[ \sum_{\tau, \tau^\prime} G_{}^{\tau} G_{}^{\tau^\prime} W_{T}^{\tau \tau^\prime}(y) \right]F^2(Q^2),
\end{equation}
where we use $G^{N}\equiv g_\chi g_N/m_\phi^2$ and the usual substitution in terms of isospin follows. For sub-GeV mediator masses, it is expected that the event yields at larger nuclear-recoil energies would be reduced when compared with the contact interaction approximation. However, they would match at smaller recoil energies. In Fig.\,\ref{fig:rate}, we show the event spectrum for $m_\phi = 300$\,MeV which gives a reasonable event rate in the region of interest for the isoscalar case.

\section{Summary and Discussions}
\label{sec:summary}
In this work we have investigated whether cosmic-ray boosted dark matter with a spin-dependent, hence momentum-dependent, interaction can produce the $248~\mathrm{keV}$ nuclear recoil reported by LZ. We have assumed a pseudo-scalar interaction as the benchmark and followed the flux from production by cosmic-ray protons, through the rock above the detector, to the recoil spectrum in xenon. The DM-nucleus cross-section was computed in three ways: the non-relativistic EFT, a FF approach, and a SR approach that keeps the relativistic DM current and uses the shell-model nuclear responses. We have found that the contact-interaction approach turns out to be inconsistent at the couplings that generate approximately one event in the region of interest as the energy-scales required turns out to invalidate the effective-operator approach. Hence we have also considered dark matter scattering through a sub-GeV pseudo-scalar mediator of mass $m_\phi$ in all three steps.

In a heavy nucleus like xenon, the NREFT and the SR cross-sections agree to an indistinguishable accuracy, while the thin-shell form factor has a zero at $250~\mathrm{keV}$ and underestimates the rate in the event window by more than an order of magnitude. The nuclear recoil rates for NREFT and SR are slightly different because the energy loss in the Earth, which is almost entirely on hydrogen, results in a larger stopping power in the NREFT approach than the SR approach. We have found that the recoil spectrum has the right shape that can explain the LZ event while being consistent with non-observations at lower energies. 

If the event continues to be statistically significant and is confirmed by other experiments, the cosmic ray boosted DM with spin-dependent interactions is a promising candidate. There are several avenues to improve the estimates provided in this paper. To begin with, the attenuation was treated using the CSDA and only for particles arriving from above. A collision-by-collision transport calculation is required, including the quasi-elastic and inelastic channels. On the model side, the detailed UV completion needs to be confronted with light meson-decay and limits from other terrestrial experiments. Other operators with a weaker momentum dependence and a wider range of DM masses are natural next steps.

\section*{Acknowledgments}
The authors would like to thank Anirban Das for useful discussions. BC is supported by an OPERA grant and seed grant (NFSG/PIL/2023/P3821) from BITS Pilani. MS acknowledges support from the Early Career Research Grant by Anusandhan National Research Foundation (ANRF/ECRG/2024/000522/PMS). BC acknowledges the use of computational resources provided by DIST-FIST Project No. SR/FST/PS-1/2017/30. We acknowledge the use of LLM for code development.

\bibliography{crdmlz_references.bib}

\end{document}